# Comparison of an OGS/Polystyrene scintillator (BSO-406) with pure OGS (BSO-100), EJ-276, EJ-309, and M600 scintillators.

M. Grodzicka- Kobylka [a,1], T. Szczesniak [a], L. Adamowski [a,b], L. Swiderski [a], K. Brylew [a], A. Syntfeld-Każuch [a], W.K. Warburton [c], J.S. Carlson [c], J.J. Valiente-Dobón [d,e],

[a] *National Centre for Nuclear Research, A. Soltana 7, PL 05-400 Świerk-Otwock, Poland*
[b] *Faculty of Physics, Warsaw University of Technology, Warsaw, Poland*
[c] *Blueshift Optics LLC, 2744 E 11th St., Ste H2, Oakland, CA 94601-1443, United States of America*
[d] *INFN, Laboratori Nazionali di Legnaro, I-35020 Legnaro, Italy*
[e] *Instituto de Física Corpuscular, CSIC-Universidad de Valencia, 46980, Valencia, Spain*

*E-mail*: `Martyna.Grodzicka@ncbj.gov.pl`

*Abstract* — In this paper, we present an investigation into the scintillation properties and pulse shape discrimination (PSD) performance of the new BSO-406, which is a blend of 40% organic glass scintillator and 60% polystyrene. We tested a cylindrical sample with dimensions of 2×2 inches. The study includes measurements of neutron-gamma discrimination capability, emission spectra, photoelectron yield, and the analysis of light pulse shapes originating from events related to gamma-rays and fast neutrons. The results were compared to data previously recorded using a pure Organic Glass Scintillator (BSO-100), an EJ-309 liquid scintillator, and EJ-276 and M600 polyurethane-based plastic scintillators.

## I. Introduction

Over the past decade or so, we have observed significant development in scintillators capable of discriminating neutron-gamma radiation [1-5]. Particular emphasis has been placed on the development of so-called safe scintillators, which are not characterized by toxicity and flammability, as well as fragility as the dimensions increase, or cloudiness under external conditions.

In this study, we present a comprehensive characterization of a novel scintillator, BSO-406, developed by Blueshift Optics. This material is a composite of 40% organic glass scintillator (OGS) and 60% polystyrene, combining the unique properties of both components. BSO-406 offers several advantages, including enhanced durability and resilience, eliminating the need for protective coatings, which often complicate handling and deployment. These features make it particularly well-suited for applications requiring intricate geometries.

The primary objective of this research was to evaluate the performance and properties of the latest commercially available BSO-406 scintillator. Its characteristics were systematically compared to those of several well-established alternatives, including pure organic glass scintillator (BSO-100, also known as OGS [1]), PSD-capable plastics such as EJ-276 [2, 6] and M-600 [5], and the liquid scintillator EJ-309 [7]. By providing a thorough comparison, this study highlights the potential of BSO-406 as a viable option for advanced radiation detection systems.

## II. Experimental details

### A. Scintillators and photomultipliers

The BSO-406, OGS glass scintillators, and plastic scintillators EJ-276 and M-600 used in the tests were polished on all surfaces and wrapped with Teflon tape on all sides, except for those surfaces that were kept transparent to transfer light to the photomultiplier's photocathode. The EJ-309 liquid scintillator was housed in a typical aluminum cylindrical container. The inner sides of the light-tight capsule were lined with a white reflector. The container was sealed with a glass window on one side. The list of the tested scintillators is presented in Table I.

The bare optical surfaces of the tested scintillators were coupled to the commercially available Hamamatsu R6233-100 spectrometry photomultiplier tube (PMT) using silicone grease (Baysilone Öl M 600 000 cSt). The R6233-100 PMT has a 76 mm diameter photocathode and is characterized by a high photocathode blue sensitivity of 15.4 µA/lmF resulting in a maximum quantum efficiency of 41% at 380 nm [8].

[1] Corresponding author.

TABLE I
MAIN PROPERTIES OF SCINTILLATORS USED IN THE STUDIES

| Crystal | Size | Shape | Peak emission [nm] | Type of scintillator | Manufacturer |
|---|---|---|---|---|---|
| BSO-406 | 2"x2" | Cylinder | 430 | Mixture (glass+plastic) | Blueshift Optics |
| OGS (BSO-100) | | | 428 | glass | Sandia National Laboratories |
| EJ-276 | | | 425 | plastic | Eljen Technologies |
| EJ-309 | | | 424 | liquid | Eljen Technologies |
| M-600 | | | 430 | plastic | Target Systemelektronik |

*B. Emission spectra*

For the measurement of emission and excitation spectra, a xenon light source (model Y1603, CVI Laser Spectral Products) was used in conjunction with a monochromator (model CM110, CVI Laser Spectral Products) featuring 0.3 mm slits for precise wavelength selection. The monochromator was fully automated and controlled via a PC through an RS232 connection. Light emerging from the slit was collimated and focused onto the sample, while the resulting emission was captured and directed through an optical fiber to a calibrated Stellarnet SilverNova spectrometer, offering a resolution of 1 nm and equipped with a 25 µm slit. The experimental setup is shown in Fig. 1.

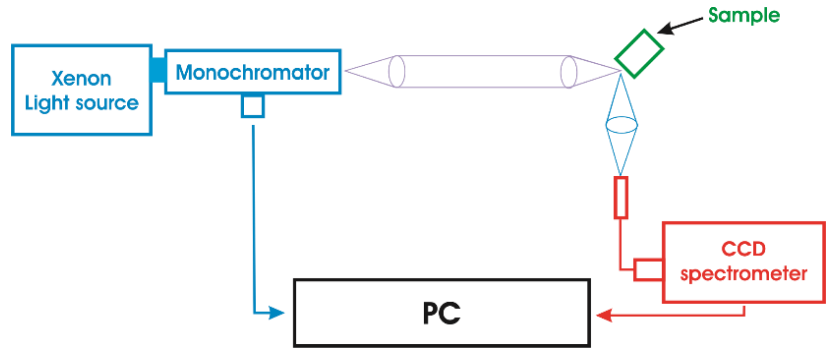

Fig. 1. Schematic diagram of the experimental setup employed for recording emission spectra.

*C. Photoelectron yield*

The number of photoelectrons was determined following the method of Bertolaccini et al. [9], which compares the scintillation peak centroid from a specific gamma-ray energy to that of a single photoelectron. None of the tested scintillators exhibit full-energy peaks beyond a certain threshold, as they lack the capability to absorb the entire energy of the γ-ray. This limitation arises from the low density and atomic number of light-material-based scintillators. Therefore, energy calibration is performed by recording the Compton continuum of gamma-rays (as demonstrated using a $^{137}$Cs source in this study), identifying the Compton edge (taken at the point where the spectrum reaches 80% of the maximum height of the Compton distribution [10-11]), and obtaining the single photoelectron spectrum from the PMT.

Light output is the ratio between the number of recorded photoelectrons, and integral quantum efficiency (IQE), where the IQE is an average of QE weighted by the emission spectrum [12].

*D. Pulse shape discrimination*

Neutron and γ-ray discrimination is typically achieved through pulse shape discrimination (PSD), which analyzes the distinct shapes of scintillation pulses they generate in detectors. In this study, the PMT output signal was processed using a DT 5730 waveform digitizer with a 500 MHz sampling rate (2 ns per sample), 14-bit resolution, and a 2 V dynamic range. A PuBe or AmBe source was used for PSD analysis, and Fig. 2 shows the experimental setup schematically.

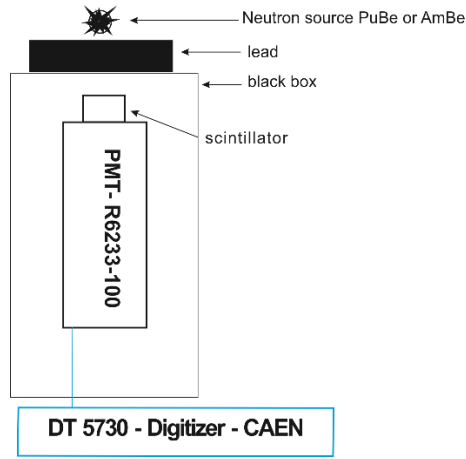

**Fig. 2.** Experimental setup for the PSD measurements.

In this work, the Charge Comparison Method (CCM), a classic PSD technique, was employed. This method relies on comparing two integrals of the current signal. The PSD parameter is defined as the ratio of the difference between the integrals of the long gate and the short gate, divided by the integral of the long gate: ($Q_{long\ gate}$ - $Q_{short\ gate}$) / $Q_{long\ gate}$. The approach is illustrated in Fig. 3, and the formula was implemented using the CAEN DT5730 digitizer [13].

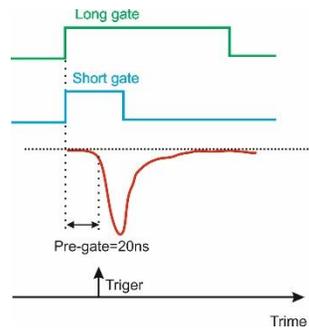

**Fig. 3.** Diagram of the n/γ discrimination method implemented in a CAEN DT5730 digitizer. The pre-gate of 20 ns contributes to the length of each gate.

To evaluate the PSD performance of the detectors, the figure of merit (FOM) was calculated as follows:

$$FOM = \frac{(peak\ separation)}{(FWHM_\gamma)+(FWHM_n)} \quad (1)$$

where FWHM represents the full width at half maximum of the peaks corresponding to neutron and γ-ray detection, projected onto the PSD parameter axis. The FOM for each pulse was assessed in narrow energy ranges (E ± 10%E) centered at 100 keVee, 300 keVee, 500 keVee, and 1000 keVee. Additionally, the FOM was also calculated for a broader energy range between 100 keVee and 1000 keVee. Energy calibration for the 2D plots was carried out using γ-ray sources $^{241}$Am, $^{137}$Cs, and $^{22}$Na, ensuring precision for selecting energy window cuts.

### E.  Light pulse shapes from gamma-rays and fast neutrons

A detailed schematic of the slow-fast experimental setup used to capture scintillation decay profiles is shown in Fig. 4. The light pulse shapes were recorded using a modified Bollinger-Thomas single-photon method [14], adapted to study pulse shapes separately for gamma-rays and fast neutrons [15]. Two photomultipliers—Photonis XP20D0 (coupled to the BSO scintillator being tested) and Hamamatsu R5320—were employed to detect single photons emitted from the sample. The PMTs were positioned on opposite sides of a light-tight tube, with only the sides of the scintillator wrapped in Teflon tape, leaving the top and bottom open to allow photon detection by both PMTs. The low time jitter of the Hamamatsu R5320, at 140 ps [16], provided the necessary time resolution for accurately capturing the scintillation decay profile, especially its fast (nanosecond) component. The method used for pulse recording is described in detail in [15]. The scintillators were irradiated with a PuBe neutron source, and a 10 cm lead brick was placed between the source and the scintillator to reduce gamma-ray interference, which could otherwise produce Cherenkov photons in the PMT window. The recorded time distribution of single photons detected by the R5320 PMT accurately reflects the scintillation light pulse shape.

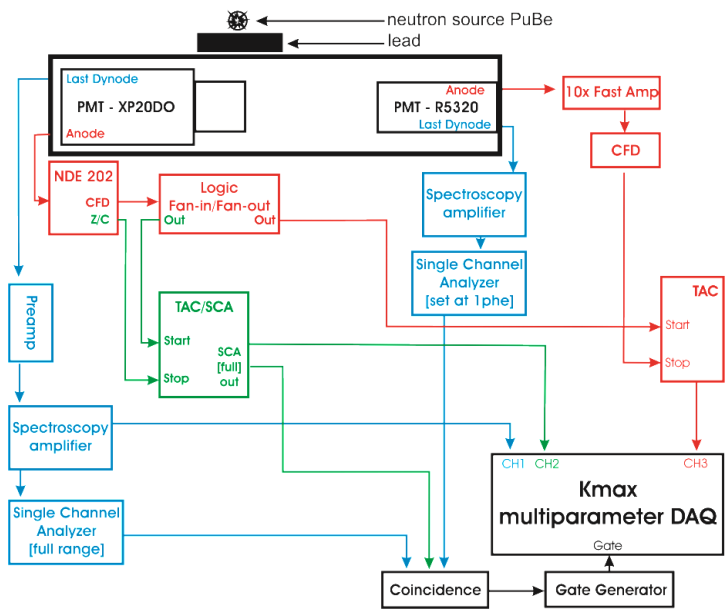

**Fig. 4.** Experimental setup for the light shape measurements using the Bollinger-Thomas single photon method with PSD, following Ref. [15].

III. RESULTS

A. *Emission spectra*

Emission spectra were recorded for the tested samples across a range of wavelengths, with excitation starting at 190 nm and proceeding to 800 nm in 1 nm steps. The spectra were corrected for spectrometer background noise. As an example, the spectrum obtained from the OGS scintillator is shown in Fig. 5. A bright diagonal line represents the excitation wavelength selected by the monochromator, while the bright region near 430 nm corresponds to the sample's emission. For the OGS sample, no emissions were observed above 580 nm. To analyze the data, vertical and horizontal cuts were taken through the emission maximum of each sample, revealing the excitation and emission spectra of these peaks.

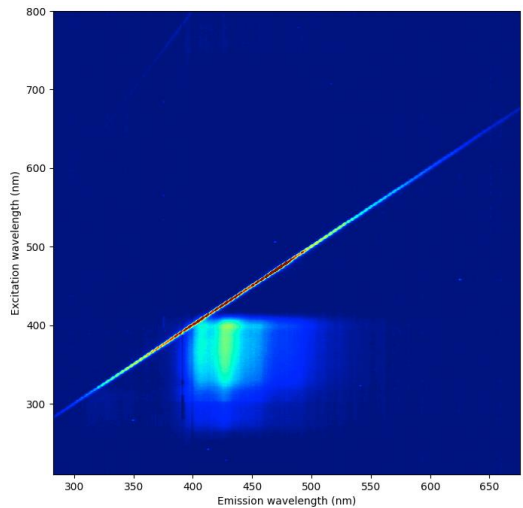

**Fig. 5.** Emission as a function of the excitation wavelength of OGS scintillator.

The observed emissions correspond to typical visible transitions from π-electron excited singlet states to the ground state in aromatic molecules [17]. However, the emission peaks of specific aromatic compounds, such as PPO, DPA, and other commonly used fluorescent components, are shifted to around 430 nm due to the incorporation of wavelength shifters during the manufacturing process of plastic scintillators [18, 19]. The primary distinction lies in the excitation spectrum, reflecting variations in the composition of the scintillator medium.

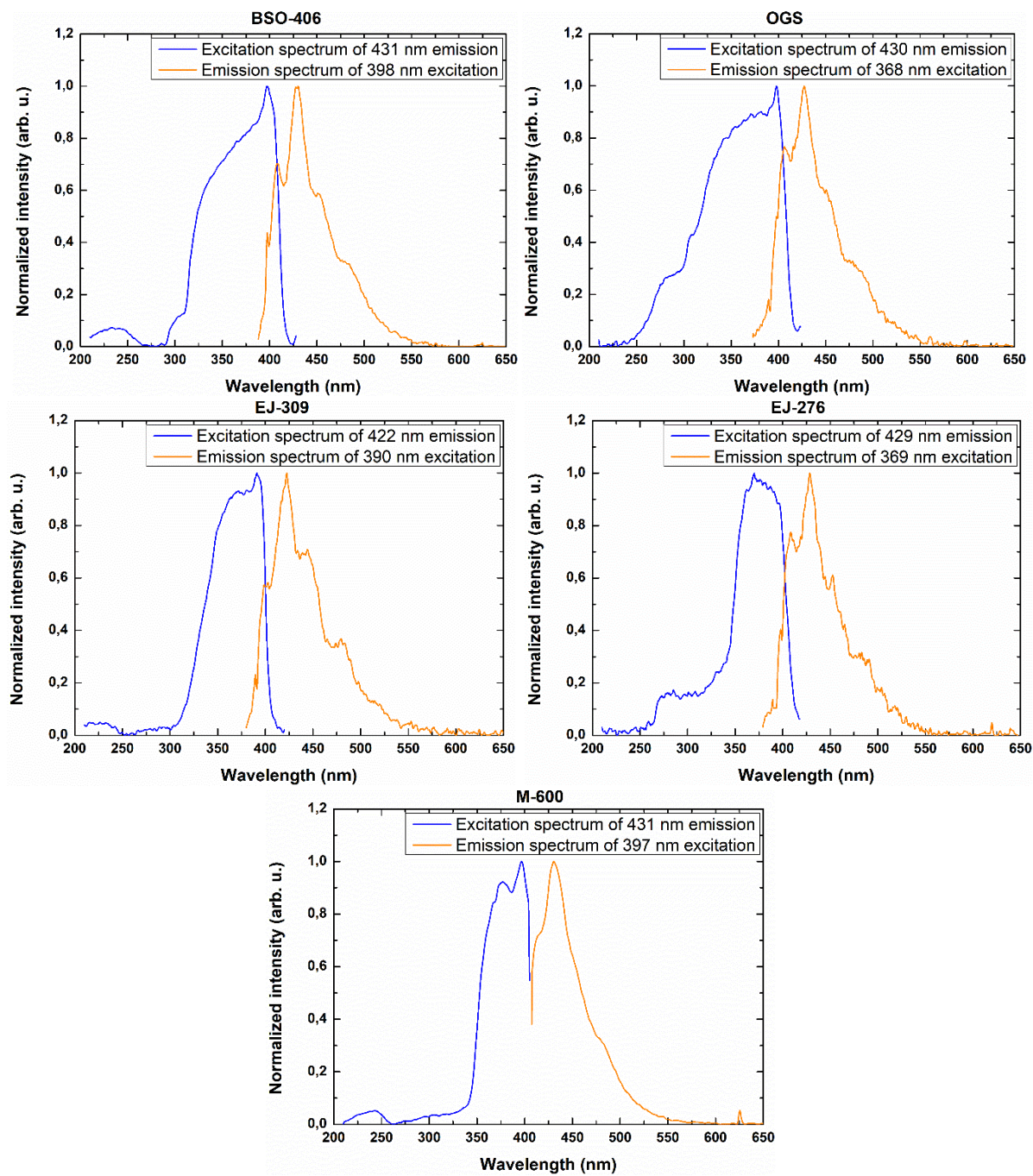

**Fig. 6.** Emission spectra of BSO-406, OGS, EJ-309, EJ-276 and M-600, respectively.

To quantitatively compare the scintillation response of the tested samples, the Internal Quantum Efficiency (IQE) was calculated for all detectors coupled with the calibrated R6233-100 PMT. The quantum efficiency (QE) of the R6233-100 is shown in Fig. 7. IQE represents a measure of the spectral matching between the scintillator and photodetector, allowing the conversion of the measured number of photoelectrons into the number of photons emitted by the sample and collected at the photocathode. According to Fig. 7, in the 340–410 nm range, the QE of the photomultiplier reaches its maximum value of 41%. Within this range, none of the tested scintillators exhibit a peak emission maximum. The closest to this range is the emission maximum of the EJ-309 scintillator, which is at 422 nm. However, the emission maxima of the remaining scintillators are similarly close, specifically at 429 nm, 430 nm, 431 nm, and 431 nm for EJ-276, OGS, BSO-406, and M600, respectively. The differences in IQE values among all the tested detectors are minimal, suggesting that the photodetector is equally well-suited to all the tested scintillators. Therefore, differences in neutron/gamma discrimination performance can be attributed solely to the intrinsic properties of the scintillators, particularly their decay profiles, scintillation yield, and emission wavelengths.

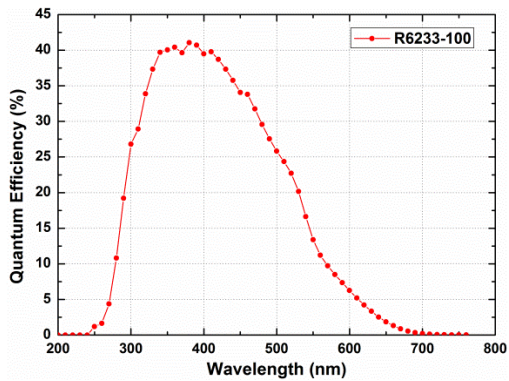

**Fig. 7.** Quantum Efficiency of the R6233-100 PMT.

*B. Photoelectron yield*

Table III presents the number of photoelectrons per MeV and the light output (LO) for all tested samples. The OGS scintillator exhibits the highest number of photoelectrons per MeV and the highest LO among the tested scintillators. The BSO-406 scintillator, a composite of 40% OGS and 60% polystyrene, produces 40% less light than pure OGS. However, its light output is comparable to that of the EJ-309 liquid scintillator. Similarly, the plastic scintillators M-600 and EJ-276 show lower light output, with comparable light output relative to each other.

Table III

Photoelectron yield and light output of investigated detectors.

| Scintillators | diameter x height (inch) | Phe number (phe/MeV) | LO (Ph/MeV) |
|---|---|---|---|
| OGS [a] | 2"x2" | 6960 ± 200 | 20 900 ± 2 000 |
| BSO-406 | 2"x2" | 4240 ± 130 | 12 700± 1 100 |
| EJ-309 [b] | 2"x2" | 4100 ± 130 | 13 200± 1 100 |
| M600 [c] | 2"x2" | 2500 ± 250 | 7 300± 1 000 |
| EJ-276 (sample from 2017) [b] | 2"x2" | 2450 | 8 200 ± 1 000 |

[a] Ref.[1], [b] Ref.[2], *(10 500±1000) ph/MeV resulting in (3140 ± 150) phe/MeV as measured in 2017* [c] Ref.[5].

*C. Pulse shapes*

The Thomas-Bollinger method was employed to record scintillation pulse shapes in a mixed field of fast neutrons and γ-rays. To distinguish the responses of the tested scintillators to the two types of radiation, PSD and energy range were used to tag events associated with specific scintillation pulses. The light pulses recorded using the BSO-406 detector, after applying constraints on energy and the PSD parameter, are shown in Fig. 8.

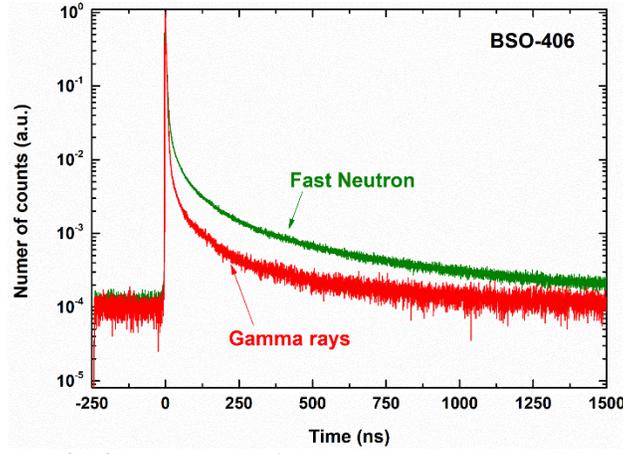

**Fig. 8.** Normalized light pulse shapes for fast neutrons and γ-rays recorded with the BSO-406 scintillator. The TAC range was set to 2,000 ns, and the energy threshold was above 300 keVee.

As shown in Fig. 8, the BSO-406 sample displays distinct decay profiles for γ-rays and fast neutrons. This characteristics enables the use of pulse shape discrimination (PSD) to differentiate fast neutrons from γ-rays. To compare the decay times of the BSO-406 scintillator with previous measurements of OGS, EJ-276, EJ-309, and M-600 scintillators under the same conditions, the light pulse components were fitted to the sum of four exponential curves using the OriginPro 8.6 software [20].

$$y = y_0 + A_1 exp(-x/\tau_1) + A_2 exp(-x/\tau_2) + A_3 exp(-x/\tau_3) + A_4 exp(-x/\tau_4) \qquad (2)$$

where: $A_1, A_2, A_3$ and $A_4$ are the amplitudes of the exponential functions, $\tau_1, \tau_2, \tau_3$ and $\tau_4$ are the scintillation decay time constants, respectively, and $y_0$ is the baseline offset originating from random coincidences.

The contribution of random coincidences was determined from the time range in the TAC spectrum preceding the light pulse's leading edge. Four exponential components were observed in the BSO-406 scintillator, corresponding to fast, medium, and two long decay components. For the OGS, EJ-309, EJ-276 and M-600 detectors, three or four exponential components were required to match the scintillation decay profiles, (see Table IV). Figure 9 shows the light pulse shapes for γ-rays and fast neutrons, alongside with multi-exponential fits for the tested BSO-406 scintillator. Each plot presents experimental data obtained using the Thomas-Bollinger method, supplemented by information from PSD. The PSD data facilitated the setting of coincidence windows based on list mode analysis, enabling the selection of events induced exclusively by fast neutrons or gamma rays (details provided in Ref. [2]). The individual exponential components fitted to each scintillation decay profile are also shown. Scintillation decay data for OGS, EJ-309, EJ-276, and M-600 crystals were acquired using the same experimental method and a nearly identical setup, as detailed in [1, 2, 5, 15]. For both the OGS and M600 scintillators, fitting was performed using only three components. However, in the case of OGS, the second long component was not observed, while for M600, the fast component was absent. It is worth noting that the first two components, fast and medium, are identical for both OGS and BSO, with the only significant difference being the intensity of the fast component. Differences in the pulse shape are mainly observed in the long component. For BSO-406, we observe four components, while for OGS, only three are present.

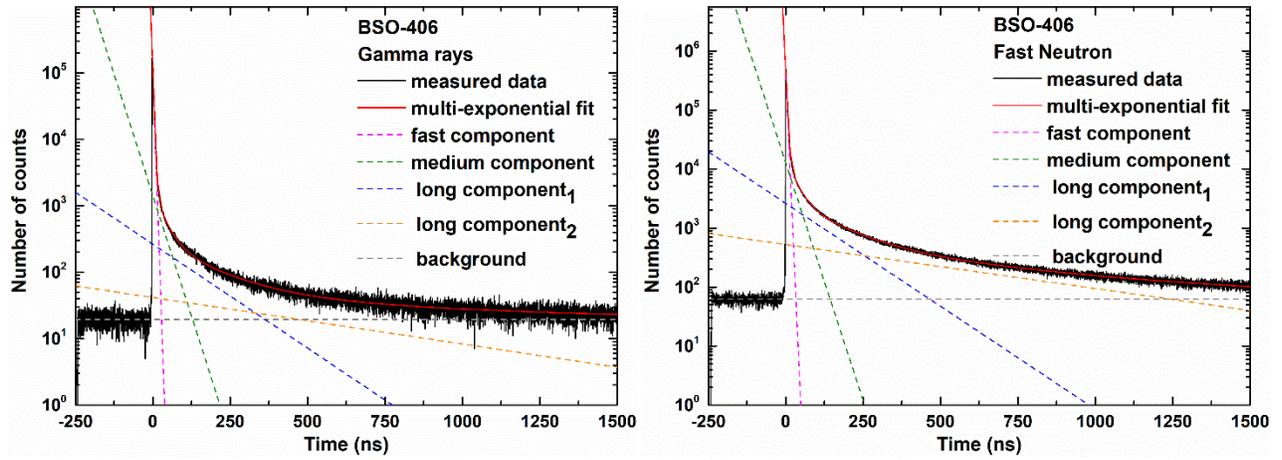

**Fig. 9.** Shows the light pulse shapes recorded for the tested BSO-406 scintillator (γ-rays on the left, fast neutrons on the right). The solid black lines represent the experimental data, while the solid red line represents the multi-exponential fit. The dashed lines correspond to the fitted decay components.

TABLE IV
FITTING PARAMETERS OF ALL COMPONENTS CALCULATED FOR THE TESTED ORGANIC SCINTILLATORS

| scintillator | radiation | Fast component Decay const. [ns] | Intensity [e)] % | Medium component Decay const. [ns] | Intensity [e)] % | Long component 1 Decay const. [ns] | Intensity [e)] % | Long component 2 Decay const. [ns] | Intensity [e)] % | Intensity 2 LONG components |
|---|---|---|---|---|---|---|---|---|---|---|
| OGS [a)] | Gamma | 3.3±0.3 | 89 | 29 ±3 | 8 | 158±20 | 3 | - | - | 3 |
| | Fast N | 3.7±0.3 | 70 | 27 ±3 | 15 | 100±10 | 16 | - | - | 16 |
| BSO-406 | Gamma | 3,3 | 78 | 29 | 9 | 138 | 8 | 620 | 5 | 13 |
| | Fast N | 3,7 | 63 | 27 | 13 | 124 | 12 | 582 | 11 | 23 |
| EJ-309 [b)] | Gamma | 3.7±0.4 | 80 | 31±3 | 10 | 140±10 | 7 | 790±80 | 3 | 10 |
| | Fast N | 4.8±0.5 | 46 | 32±3 | 24 | 140±10 | 20 | 620±60 | 11 | 31 |
| EJ-276 [c)] | Gamma | 4.0±0.4 | 70 | 16±2 | 12 | 98±10 | 8 | 690±70 | 8 | 16 |
| | Fast N | 3.9±0.4 | 47 | 18±2 | 13 | 106±10 | 13 | 800±80 | 27 | 30 |
| M-600 [d)] | Gamma | - | - | 13.9±0.1 | 85 | 93±2 | 6 | 532±5 | 9 | 15 |
| | Fast N | - | - | 14.1±0.1 | 68 | 82±1 | 13 | 501±19 | 19 | 32 |

[a)] from [1], [b)] from [15], [c)] from [2], [d)] from [5], [e)] $Intensity = \frac{A_i \times \tau_i}{\sum A_i \times \tau_i}$,

The ability of scintillators to discriminate between neutrons and γ-rays is defined by the ratio of slow component intensities induced by neutrons (recoil protons) to those induced by γ-rays (primary electrons). The data in Table IV allow the calculation of these ratios for all tested scintillators (OGS, BSO-406, EJ-309, EJ-276, and M600), with the results summarized in Table V. Considering the ratio of the combined intensities of the medium and first long components (ranging from 13 ns to 200 ns) for fast neutrons and γ-rays, the respective values are 2.8, 1.5, 2.6, 1.3, and 0.9 for OGS, BSO-406, EJ-309, EJ-276, and M600. A higher ratio indicates better PSD performance for the scintillator. These results position OGS and EJ-309 on par, followed by BSO-406, EJ-276, and lastly, M-600, with a ratio of 0.9.

Table V
The ratio of neutron to gamma-ray component intensities

| scintillator | radiation | Intensity of medium and first long components Intensity [c)] % | Neutron/Gamma Ratio |
|---|---|---|---|
| OGS | Gamma | 11 | 2.8 |
| | Fast N | 31 | |
| BSO-406 | Gamma | 17 | 1.5 |
| | Fast N | 25 | |
| EJ-309 [a)] | Gamma | 17 | 2.6 |

| | Fast N | 44 | |
|---|---|---|---|
| EJ-276[b)] | Gamma | 20 | 1.3 |
| | Fast N | 26 | |
| M-600 | Gamma | 91 | 0.9 |
| | Fast N | 81 | |

*D. PSD performance*

In this part of the study, we compare the n/γ discrimination performance of the BSO-406 scintillator with that of OGS, EJ-309, and the plastic scintillators EJ-276 and M600. Each scintillator was characterized following its individual optimization procedure. The optimization involved determining the short and long gate lengths that yielded the highest FOM value for the tested scintillator. The short gate length was adjusted in 10 ns increments from 40 ns to 100 ns, with two additional measurements taken within the range indicating the best result. Similarly, the long gate length was varied in 100 ns steps from 400 ns to 1000 ns, also with two additional measurements. The pre-gate was set to 20 ns (see Fig. 3). The optimization results for the BSO-406 scintillator are shown in Fig. 10. The best FOM was obtained with a short gate of 66 ns and a long gate of 600 ns. It is worth noting that selecting the optimal short gate is crucial, as this value has a greater impact on the final FoM result. The FOM values for energy cuts at 100, 300, 500, and 1000 keVee are 1.3, 2.23, 2.46, and 2.59, respectively.

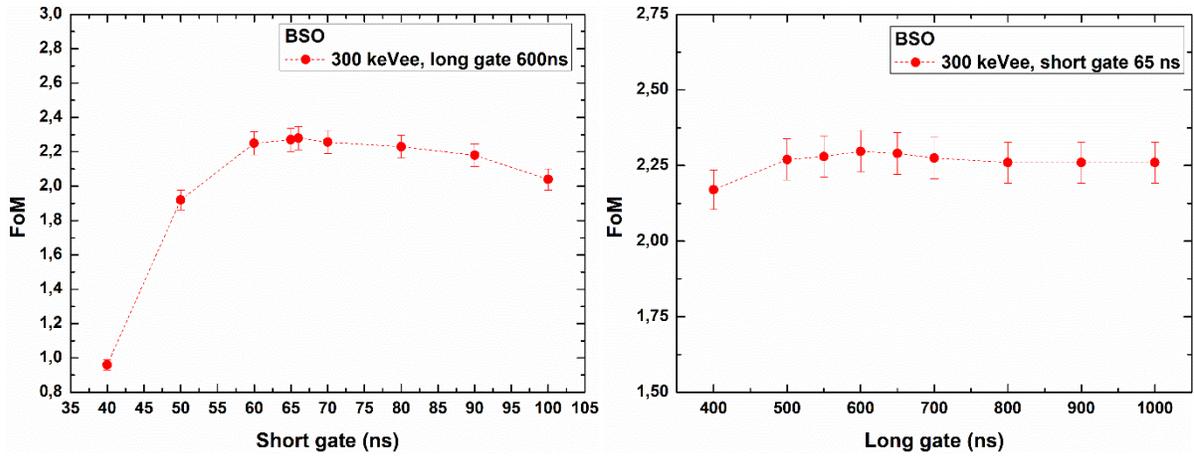

**Fig. 10**. The calculated FoM values as a function of the short and long integration gates for the 300 keVee energy window.

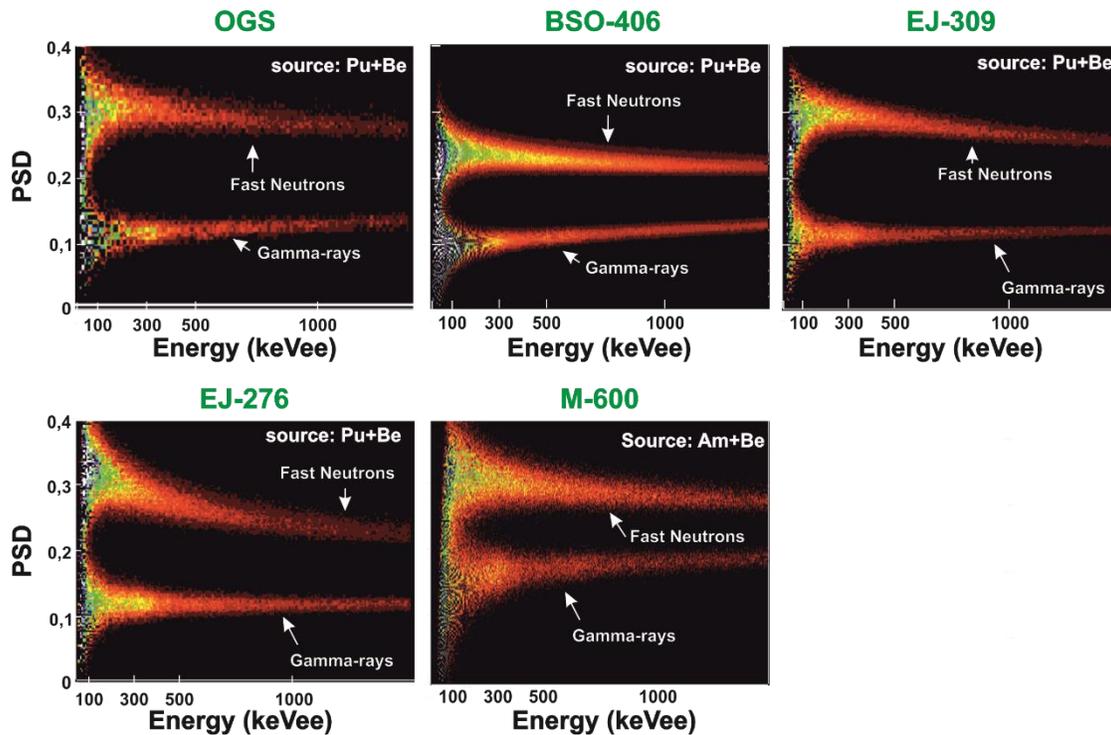

**Fig.11.** 2D histograms of PSD versus energy recorded with PuBe sources for BSO-406, OGS, EJ-309, EJ-276 and M-600.

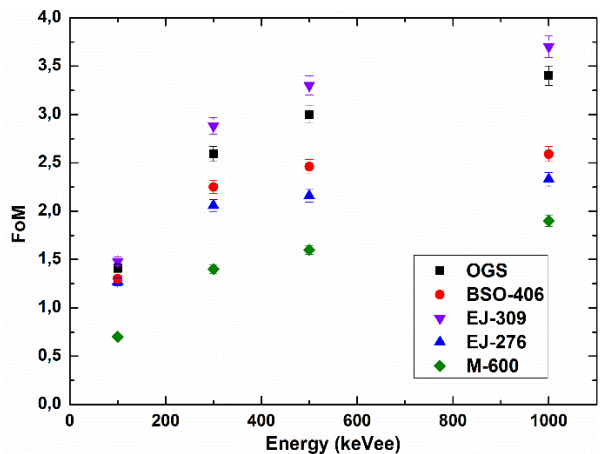

**Fig. 12.** Comparison of the Figure of Merit (FOM) for 2"×2" scintillator samples: BSO-406, OGS, EJ-276, EJ-309, and M600.

The pulse shape discrimination (PSD) capabilities of the BSO-406 sample were compared with the n/γ discrimination performance of the OGS, EJ-276, EJ-309, and M-600 scintillators (see Fig. 11). All tested scintillators demonstrated clear separation of fast neutron and gamma-ray events starting at an energy of 100 keVee. The figure of merit (FOM) values, calculated for energies of 100, 300, 500, and 1000 keVee within ranges of ±0.1 of the energy, as well as across an energy window of 100–1000 keVee, indicate that the liquid scintillator EJ-309 achieves slightly better performance compared to the other scintillators (see Table VI). The performance assessed through the evaluation of FOM results is consistent with the recorded scintillation yields and the ratios of the first long components measured for neutrons and gamma rays. It is noteworthy that the performance of the OGS scintillator is nearly on par with the liquid scintillator EJ-309, making OGS a viable alternative for fast neutron detection, especially in challenging environmental conditions where solid detectors are preferred over liquid or gas-based detectors. Meanwhile, the BSO-406 scintillator proves to be an attractive alternative to the OGS scintillator due to its lack of need for coatings and better robustness. It is also worth noting that the results obtained for BSO-406 are the best ever recorded for plastic scintillators, taking into account that 60% of its composition is polystyrene.

**Table VI**

FOM for energy cuts at 100, 300, 500 and 1000 keVee and an energy window between 100-1000 keVee measured with EJ-309, EJ-276, stilbene single crystal and OGS scintillators.

| Scintillator | OGS | BSO-406 | EJ-309 | EJ-276 | M-600 |
|---|---|---|---|---|---|
| Short gate  Long gate | 70  400 | 65  600 | 60  800 | 74  700 | 70  800 |
| 100 keVee | 1.41±0.04 | 1.3±0.04 | 1.48±0.04 | 1.27±0.03 | 0.7±0.02 |
| 300 keVee | 2.59±0.07 | 2.23±0.07 | 2.88±0.08 | 2.06±0.06 | 1.4±0.04 |
| 500 keVee | 3.0±0.1 | 2.46±0.07 | 3.3±0.1 | 2.16±0.06 | 1.6±0.05 |
| 1000 keVee | 3.4±0.1 | 2.59±0.07 | 3.7±0.1 | 2.33±0.07 | 1.9±0.06 |
| 100-1000keVee | 2.17±0.06 | 1.71±0.08 | 2.34±0.07 | 1.32±0.04 | - |
| Nphe at 300 keV | 2020±100 | 1280 | 1230±60 | 740±40 | 750±40 |

## IV. CONCLUSIONS

The new BSO-406 scintillator, developed by Blueshift Optics, is a blend of 40% OGS scintillator and 60% polystyrene. Its light output is reduced by 40% compared to pure OGS, but remains relatively high, comparable to the EJ-309 scintillator, at around 13,000 ph/MeV. It features an emission spectrum peaking at approximately 430 nm, making it compatible with most available photodetectors. While the pulse decay of BSO-406 is slower than that of OGS, it is still fairly fast compared to other scintillators on the market. The PSD performance, though slightly lower than OGS, is the best among plastic scintillators.

Considering the practical advantages of organic glass scintillators, such as better robustness, non-toxicity, and non-flammability, both OGS and BSO-406 are viable alternatives to liquid scintillators for neutron detection with gamma discrimination. This makes them particularly suitable for applications like handheld neutron/gamma ray detectors used in real-world scenarios.

### Acknowledgements
This work was supported in part by the European Union (ChETEC-INFRA, project no. 101008324).